\documentclass[12pt]{iopart}
\expandafter\let\csname equation*\endcsname\relax
\expandafter\let\csname endequation*\endcsname\relax
\usepackage{amsmath}
\usepackage{amssymb}
\usepackage{geometry}
\usepackage{cite}
\begin{document}

\hfill UPR-1268-T

\title[An Analysis of the Wave Equation for the $U(1)^{2}$ Gauged ...]{An Analysis of the Wave Equation for the $U(1)^{2}$ Gauged Supergravity Black Hole}

\author{Tolga Birkandan$^{1}$ \& Mirjam Cveti\v{c}$^{1,2,3}$}

\address{$^1$ Department of Physics, Istanbul Technical University, Istanbul 34469, Turkey.}
\address{$^2$ Department of Physics and Astronomy, University of Pennsylvania, Philadelphia, PA 19104, USA.}
\address{$^3$ Center for Applied Mathematics and Theoretical Physics, University of Maribor, Maribor, Slovenia.}
\eads{\mailto{birkandant@itu.edu.tr}, \mailto{cvetic@hep.upenn.edu}}

\begin{abstract}
We study the massless Klein-Gordon equation in the background of the most general rotating dyonic AdS black hole in gauged $\mathcal{N}=2$, $U(1)^{2}$ gauged supergravity in $D=4$, given by Chow and Comp\`{e}re [Phys. Rev. D\textbf{89} (2014) 065003]. The angular part of the separable wave equation is of the Heun type, while the radial part is a Fuchsian equation with five regular singularities. The radial equation is further analyzed and written in a specific form, that reveals the pole structure of the horizon equation, whose residua are expressed in terms of the surface gravities and angular velocities associated with respective horizons. The near-horizon (near-)extremal limits of the solution are also studied, where the expected  hidden conformal symmetry is revealed. Furthermore, we present the retarded Green's functions for these limiting cases. We also comment on the generality of the charge-dependent parts of the metric parameters and address some further examples of limiting cases.
\end{abstract}


%\submitto{\CQG}
\maketitle
%%%%%%%%%%%%%%%%%%%%%%%%%%%%%%%%%%%%%%%%%%%%%%%
\section{Introduction}
%%%%%%%%%%%%%%%%%%%%%%%%%%%%%%%%%%%%%%%%%%%%%%%
Klein-Gordon equation is an essential part of the studies on the internal structure of black holes. Separating and solving the scalar equation, mainly its radial part, is the starting point of many studies including scattering calculations. Moreover, the analysis of the symmetries of the radial equation provides insights into gauge/gravity correspondence.

It is intriguing that for all the known black hole solutions of ungauged and gauged supergravity theories in four and five dimensions, the massless Klein Gordon wave equation is separable. This feature signifies the existence of the Killing-St\"ackel tensor. For a  recent review on the subject, see \cite{Chow:2008fe} and references therein.

Early on, for the general rotating multi-charged black holes of maximally supersymmetric ungauged supergravities where the horizon equation has two solutions \cite{Cvetic:1996kv,Cvetic:1996xz}, it was shown that the radial part of the scalar equation could be written in a form in which the residua of the poles associated with the two horizons could be specified in terms of the surface gravities and angular velocities at the respective horizons \cite{Cvetic:1997xv,Cvetic:1997uw}. This important property signifies the contributions of both horizons in the wave equation, along with their thermodynamic quantities in the associated residua.

This feature also  works \cite{Cvetic:2014sxa}  for the most general four-dimensional black holes of maximally supersymmetric ungauged supergravity, constructed by Chow and Comp\`ere \cite{Chow:2013tia} \footnote{The structure of the radial equation of the Kerr black hole can be related to the monodromy data \cite{Castro:2013lba,Castro:2013kea},  which carry the thermodynamic quantities  associated with this configuration and are associated with the  hidden conformal symmetry \cite{Castro:2010fd}.

In another approach, proposed in  \cite{Cvetic:2011dn,Cvetic:2011hp,Cvetic:2012tr,Cvetic:2014sxa},  the modification of the warp factor of the general black holes of the ungauged supergravity does not change the thermodynamic quantities associated with the two horizons. A specific choice of the warp factor  results in a so-called subtracted geometry whose massless Klein-Gordon equation is separable and whose radial part is solved by hypergeometric functions, which signifies the appearance of $SL(2,R)^2$ conformal symmetry, associated with the internal structure of the black hole. For further recent studies, see \cite{Cvetic:2013lfa,Cvetic:2014ina,Cvetic:2014eka} and references therein. The detailed thermodynamics of the subtracted geometry was recently studied in \cite{Cvetic:2014nta}.}.

This intriguing structure of the wave equation for  ungauged supergravity black holes can be generalized to general black holes of five-dimensional  gauged supergravity, which have multiple horizons.   In this case the  radial equation  can be written in terms of the poles associated with {\it each of the multiple  horizons} and residua are  again expressed in  terms of  a specific combination of the  surface  gravities and angular velocities   associated with each respective  horizon.  This feature of the wave equation was explicitly demonstrated  \cite{Birkandan:2011fr,Birkandan:2014cva}   for the general  black hole in minimal five-dimensional gauged supergravity with three equal charges \cite{Chong:2005hr}. Most recently,   the study of  the massless Klein-Gordon  equation for the most general black hole solution of  five-dimensional maximally supersymmetric gauged supergravity  with  $U(1)^{3}$ gauge symmetry \cite{Wu:2011gq} was  presented in   \cite{Birkandan:2014vga}. In spite of the complexity of the metric,  the separability of wave equation was demonstrated and the radial part  was cast in a form with the poles of the horizon equation, whose residua were again written in the specific form containing the associated surface gravities and angular velocities.

The horizon equation is quadratic in radial coordinate, $r$, for the black holes in five dimensions. Thus, applying a transformation in the form $u=r^2$ to the radial coordinate, one can deal effectively only with three poles in terms of the new radial coordinate $u$, instead of six poles of the horizon equation in terms of the $r$ coordinate. Therefore, only four regular singularities (three horizons and positive infinity) appear in the radial equation, which yields a solution in terms of the general Heun's functions \cite{Ronveaux95,Birkandan:2006ac,Hortacsu:2011rr}.

For all the examples above, the scalar field Ansatz is of the form
\begin{equation}
\Phi (t, r,\theta,\phi,\psi )=e^{-i\omega t}e^{im_{\phi }\phi }e^{im_{\psi }\psi}R(r)S(\theta ),
\end{equation}
and one can see that we can generally write
\begin{equation}
\frac{d}{dr}\left( X\frac{dR}{dr}\right) +\left[ \left( \overset{\#poles}{%
\underset{i=1}{\sum }}n_{i}\frac{\left( \omega -m_{\phi }\Omega _{\phi
i}-m_{\psi }\Omega _{\psi i}\right) ^{2}}{4\kappa _{i}^{2}\left(
r-r_{i}\right) }\right) +\tilde{n}\omega ^{2}+\hat{n}\omega +\breve{n}\right]
R=0,  \label{eq:fullrad}
\end{equation}
as the radial part of the massless Klein-Gordon equation. Here, $\Omega_{\phi i}$ and $\Omega _{\psi i}$ are the angular velocities, and $\kappa_{i}$ is the surface gravity associated with the horizon $r_{i}$. $X=0$ is the horizon equation and the $n_{i}$ ($i=1..\#poles$) constants are given by
\begin{equation}
n_{i}=\frac{dX}{dr}|_{r=r_{i}}.
\end{equation}
The other factors $\tilde{n}$ and $\hat{n}$ may have a radial dependence while $\breve{n}$ usually carries only the eigenvalue of the angular operator.

We should also point out another interesting property of general black holes: the product of entropies at all horizons of black holes have a universal nature in  gauged supergravities and higher derivative gravity theories, and they depend only on the quantized charges, quantized angular momenta \cite{Larsen:1997ge,Cvetic:1997vp} and cosmological constant \cite{Cvetic:2010mn,Castro:2013pqa,Cvetic:2013eda}. This feature also underscores an important role that the thermodynamics associated with {\it all} the horizons plays in studies of the internal structure of general black holes.

The goal of this paper is to address the massless Klein-Gordon wave equation for the black holes in four-dimensional gauged supergravity theories, which still requires, analogous studies as  those performed for  black holes in five-dimensional gauged supergravity theories.  In this case, one limitation  is that the most general black holes in maximally supersymmetric gauged supergravity are specified  by four electric and four magnetic charges of  $U(1)^4$ gauge symmetries, and due to lack of the generating techniques in gauged supergravity theories, this solution has not yet been obtained.

We therefore have to confine ourselves to the study  in the background of the  most general black hole in  four-dimensional  $\mathcal{N}=2$  gauged supergravity with $U(1)^2$ gauge symmetry,  given in  \cite{Chow:2013gba}. The solution describes rotating AdS black holes specified by  the mass, angular momentum, cosmological constant, and  two independent electric  and  two independent magnetic charges. This is currently the most general rotating black hole solution in four dimensional maximal supersymmetric gauged supergravity. Due to the anticipated $S$-duality symmetry, this solution has the same  metric form as the one with two independent electric-charges only, obtained in \cite{Chong:2004na}. We also need to mention the solution containing one charge parameter less than
the one in \cite{Chow:2013gba}, but having general Fayet-Iliopoulos parameters \cite{Gnecchi:2013mja}.

The main goal is to study the  radial wave equation and cast it in a form analogous to  equation (\ref{eq:fullrad}). Note however, that in four-dimensional gauged supergravity, the horizon equation is  associated with  {\it four} zeroes and thus the pole structure results in a Fuchsian equation with five regular singularities, each of the four horizons and positive infinity. Indeed, the analysis confirms  the anticipated analogous structure, which is similar to equation (\ref{eq:fullrad}). We  also  study the near-horizon (near-)extremal limits of the solution and confirm the anticipated hidden conformal symmetry there. We will finish by calculating the retarded Green's functions for these limiting cases. In the appendix, we will also comment on $S$-duality feature of the metric.

In  Section 2 we present the metric and the thermodynamic quantities of the  general $U(1)^2$ gauged supergravity black hole, summarizing the results of \cite{Chow:2013gba}. The massless Klein-Gordon wave  equation is studied in Section 3, both in its full form and in near horizon (near-)extremal form which reveals the hidden conformal symmetry.  We also present the explicit form of the retarded Green's functions there. Conclusions are given in Section 4. In  Appendix A,  we study the charge dependence explicitly and comment on the generality of the metric parametrization. In  Appendix B, we comment further on some limiting cases of the radial equation.
%%%%%%%%%%%%%%%%%%%%%%%%%%%%%%%%%%%%%%%%%%%%%%%
\section{General $U(1)^2$ gauged supergravity black hole}
%%%%%%%%%%%%%%%%%%%%%%%%%%%%%%%%%%%%%%%%%%%%%%%
The solution was given in \cite{Chow:2013gba} as
\begin{eqnarray}
ds^{2}&=&-\frac{R_{g}}{W}\left( dt-\frac{a\sin ^{2}\theta +4N_{g}\sin
^{2}\left( \frac{\theta }{2}\right) }{\Xi }d\tilde{\phi}\right) ^{2} \notag \\
&&+\frac{\Theta _{g}\sin ^{2}\theta }{W}\left( adt-\frac{L}{\Xi }d\tilde{\phi}\right)
^{2} +W\left( \frac{dr^{2}}{R_{g}}+\frac{d\theta ^{2}}{\Theta _{g}}\right) ,
\end{eqnarray}
and $\tilde{\phi}=\phi -ag^{2}t$ is defined in order to have a static coordinate frame at infinity. The functions appearing in the metric are given by
\begin{eqnarray}
R_{g} &=&r^{2}-2mr+a^{2}+e^{2}-N_{g}^{2} \notag \\
&&+g^{2}\left[r^{4}+(a^{2}+6N_{g}^{2}-2v^{2})r^{2}+3N_{g}^{2}\left( a^{2}-N_{g}^{2}\right)\right] , \\
L &=&r^{2}+(N_{g}+a)^{2}-v^{2}, \\
\Theta _{g} &=&1-a^{2}g^{2}\cos ^{2}\theta -4ag^{2}N_{g}\cos \theta
, \\
W &=&r^{2}+(N_{g}+a\cos \theta )^{2}-v^{2}, \\
\Xi &=&1-a^{2}g^{2}-4aN_{g}g^{2}.
\end{eqnarray}
The parameters of the solution are the gauge coupling constant $g$, the mass parameter $m$, the NUT (Newman-Unti-Tamburino) charge $N_g$, the rotation parameter $a$, and four charge parameters which are given explicitly in Appendix A. The full solution contains a complex scalar and two $U(1)$ field strengths, whose explicit expressions can be found in \cite{Chow:2013gba}. This metric is a general one having many subcases. We analyze the explicit charge-dependence of the parameters in Appendix A. It is seen that the special case of two electric charges that was obtained previously in \cite{Chong:2004na} is no less complicated than this general form due to $S$-duality. As a result of this property, one could have done the wave equation analysis of the solution with two electric charges and apply an analog of the $S$-duality transformation.

The horizon equation $R_{g}$ is not quadratic in $r$. Therefore we need to work with all four poles of the equation which can be written in the form
\begin{equation}
R_{g}=g^{2}(r-r_{1})(r-r_{2})(r-r_{3})(r-r_{4}),
\end{equation}
and the roots have simple relations, namely
\begin{eqnarray}
\overset{4}{\underset{i=1}{\sum }}r_{i} =0, \\
\underset{j<i=1}{\overset{4}{\sum }}r_{i}r_{j} =\frac{1}{g^{2}}%
+6N_{g}^{2}+a^{2}-2v^{2}, \\
\overset{4}{\underset{k<j<i=1}{\sum }}r_{i}r_{j}r_{k} =\frac{2m}{g^{2}}, \\
\underset{i=1}{\overset{4}{\prod }}r_{i} =\frac{%
a^{2}+e^{2}+(3a^{2}g^{2}-1)N_{g}^{2}-3N_{g}^{4}g^{2}}{g^{2}}.
\end{eqnarray}
We can use the transformation
\begin{equation}
y=a^{2}\cos ^{2}\theta ,
\end{equation}
to get rid of the trigonometric functions in the metric. This transformation
simplifies the singularity structure of the angular equation.

For the zero NUT charge ($N_{g}=0$), angular velocity, surface gravity and entropy were given as:
\begin{eqnarray}
\Omega _{i} &=&a\left( \frac{\Xi }{L(r_{i})}+g^{2}\right) , \\
\kappa _{i} &=&\frac{1}{2L(r_{i})}\left( \frac{dR_{g}}{dr}\right)
|_{r=r_{i}}=2\pi T_{i}, \\
S_{i} &=&\frac{\pi L(r_{i})}{\Xi G}.
\end{eqnarray}
respectively, in \cite{Chow:2013gba}. From this point on, we will work for $N_{g}=0$ as well. For $N_{g}=0$ and applying $y=a^{2}\cos ^{2}\theta ,$ we have
\begin{equation}
ds^{2}=-\frac{R_{g}}{W}\left( dt-\frac{a^{2}-y}{a\Xi }d\tilde{\phi}\right)
^{2}+\frac{\Theta _{g}(a^{2}-y)}{a^{2}W}\left( adt-\frac{L}{\Xi }d\tilde{\phi%
}\right) ^{2}+W\left( \frac{dr^{2}}{R_{g}}+\frac{dy^{2}}{4y(a^{2}-y)\Theta
_{g}}\right) ,
\end{equation}
we still have $\tilde{\phi}=\phi -ag^{2}t$ and the functions appearing in the metric become
\begin{eqnarray}
R_{g} &=&r^{2}-2mr+a^{2}+e^{2}+g^{2}\left[ r^{4}+(a^{2}-2v^{2})r^{2}%
\right] , \\
L &=&r^{2}+a^{2}-v^{2}, \\
\Theta _{g}&=&1-g^{2}y, \\
W &=&r^{2}+y-v^{2}, \\
\Xi &=&1-a^{2}g^{2}.
\end{eqnarray}
We will be using this form of the metric in the rest of the paper.
%%%%%%%%%%%%%%%%%%%%%%%%%%%%%%%%%%%%%%%%%%%%%%%
\section{Massless Klein-Gordon equation}
%%%%%%%%%%%%%%%%%%%%%%%%%%%%%%%%%%%%%%%%%%%%%%%
The massless Klein-Gordon equation is given as
\begin{equation}
\frac{1}{\sqrt{-g}}\partial _{\mu }\left( \sqrt{-g}g^{\mu \nu }\partial_{\nu }\varphi \right) =0.
\end{equation}
The equation is separable as $Wg^{\mu \nu }$ elements are separable, as mentioned in \cite{Chow:2013gba}. The solution can be written with the Ansatz
\begin{equation}
\varphi =e^{-i\omega t}e^{ik\phi }X(r)Y(y).
\end{equation}
%%%%%%%%%%%%%%%%%%%%%%%%%%%%%%%%%%%%%%%%%%%%%%%
\subsection{Full Klein-Gordon equation}
%%%%%%%%%%%%%%%%%%%%%%%%%%%%%%%%%%%%%%%%%%%%%%%
The angular equation reads
\begin{equation}
\frac{d^{2}Y}{dy^{2}}+\left[ {\frac{1}{\Theta _{g}}\frac{d\Theta _{g}}{dy}+%
\frac{\left( {a}^{2}-3\,y\right) }{2y\left( {a}^{2}-y\right) }}\right] \frac{%
dY}{dy}+{\frac{c_{{1}}\left( {a}^{2}-y\right) \Theta _{g}-\left[ \left( {a}%
^{2}-y\right) \left( akg^{2}-\omega \right) +ak\Xi \right] ^{2}}{4y\left( {a}%
^{2}-y\right) ^{2}\Theta _{g}^{2}}}Y=0,
\end{equation}
and its solution can be found in terms of general Heun's functions, namely
\begin{eqnarray}
\fl Y &=&C_{1}\left( {g}^{2}y-1\right) ^{\,{\frac{\omega }{2g}}}\left( y-{a}^{2}\right) ^{\frac{k}{2}}\times   \notag \\
\fl &&\left\{ H\left( {\frac{1}{{a}^{2}{g}^{2}}},\,{\frac{{a}^{2}\omega \left(g+\omega \right) +{k(k+1)}-2\,ak\omega -c_{{1}}}{4{a}^{2}{g}^{2}}},\,{\frac{gk+\omega }{2g}}\right. \right.   \notag \\
\fl &&\left. ,\,{\frac{\left( k+3\right) g+\omega }{2g}},\frac{1}{2},k+1,{\frac{y}{{a}^{2}}}\right)   \notag \\
\fl &&+C_{2}\sqrt{y}H\left( {\frac{1}{{a}^{2}{g}^{2}}},{\frac{(g+\omega)(2g+\omega ){a}^{2}+{(k+1)(k+2)}-2\,ak\omega -c_{{1}}}{4{a}^{2}{g}^{2}}}\right.   \notag \\
\fl &&\left. \left. ,{\frac{g\left( k+1\right) +\omega }{2g}},{\frac{g\left(k+4\right) +\omega }{2g}},\frac{3}{2},k+1,{\frac{y}{{a}^{2}}}\right)\right\},
\end{eqnarray}
where $C_1$, $C_2$ are arbitrary constants. Here, $c_{1}$ is the separation constant. The radial part can be written as
\begin{equation}
\frac{d}{dr}\left( R_{g}{\frac{dX}{d{r}}}\right) +\left[ \frac{\left\{
\left( {r}^{2}+{a}^{2}-{v}^{2}\right) \omega -ak\left[ 1+\left( {r}^{2}-{v}%
^{2}\right) {g}^{2}\right] \right\} ^{2}}{R_{g}}-c_{1}\right] X=0,
\end{equation}
and this equation has five regular singularities at $\{r_{1},r_{2},r_{3},r_{4},\infty \}$.

Following the existing literature we mentioned in the introduction, we can expect to have a form of the radial equation with the surface gravities and angular velocities as the residua of the poles. We can propose
\begin{equation}
\frac{d}{dr}\left( R_{g}{\frac{dX}{d{r}}}\right) +\left[ \left( \overset{4}{%
\underset{i=1}{\sum }}n_{i}\frac{\left( \omega -k\Omega _{i}\right) ^{2}}{%
4\kappa _{i}^{2}\left( r-r_{i}\right) }\right) +n_{5}\omega ^{2}+n_{6}\right]
X=0.
\end{equation}
We have the constants $n_{i}$ ($i=1,2,3,4$) associated with the poles as the derivative of the horizon equation calculated at the corresponding singularity. We expect to have
\begin{eqnarray}
n_{1} &=&\left( \frac{dR_{g}}{d{r}}\right)
|_{r=r_{1}}=g^{2}(r_{1}-r_{2})(r_{1}-r_{3})(r_{1}-r_{4}), \\
n_{2} &=&\left( \frac{dR_{g}}{d{r}}\right)
|_{r=r_{2}}=-g^{2}(r_{1}-r_{2})(r_{2}-r_{3})(r_{2}-r_{4}), \\
n_{3} &=&\left( \frac{dR_{g}}{d{r}}\right)
|_{r=r_{3}}=g^{2}(r_{1}-r_{3})(r_{2}-r_{3})(r_{3}-r_{4}), \\
n_{4} &=&\left( \frac{dR_{g}}{d{r}}\right)
|_{r=r_{4}}=-g^{2}(r_{1}-r_{4})(r_{2}-r_{4})(r_{3}-r_{4}).
\end{eqnarray}
Using these coefficients, the surface gravities and angular velocities given in the previous section, we can see that the form we proposed holds. The calculation requires a symbolic analysis of the calculated and proposed equation using the pole relations we gave in the previous section. The remaining coefficients are
\begin{eqnarray}
n_{5} &=&\frac{{a}^{2}{g}^{2}k^{2}}{{\omega }^{2}}-\frac{2ka}{{\omega }}+%
\frac{{1}}{{g}^{2}}, \\
n_{6} &=&-c_{1}.
\end{eqnarray}
Thus we finally we have
\begin{equation}
\frac{d}{dr}\left( R_{g}{\frac{dX}{d{r}}}\right) +\left[ \left( \overset{4}{%
\underset{i=1}{\sum }}n_{i}\frac{\left( \omega -k\Omega _{i}\right) ^{2}}{%
4\kappa _{i}^{2}\left( r-r_{i}\right) }\right) -2\,ak{\omega }+\frac{\omega
^{2}}{{g}^{2}}+{a}^{2}{g}^{2}k^{2}-c_{1}\right] X=0,  \label{fullradial}
\end{equation}
as the radial part of the massless Klein-Gordon equation reflecting the pole behavior of the horizon equation and the thermodynamic properties of the spacetime. This is a Fuchsian equation with five regular singularities at $\{r_{1},r_{2},r_{3},r_{4},\infty \}$.

We can think of a special case with pairwise equal poles, namely $r_{1}=r_{2} $ and $r_{3}=r_{4}$. Then using the root property $r_{1}+r_{2}+r_{3}+r_{4}=0$ we have $r_{3}=r_{4}=-r_{1}$. The horizon equation yields $R_{g}=g^{2}(r-r_{1})^{2}(r+r_{1})^{2}$ and as the surface gravity depends on the derivative of this equation calculated on the associated horizon, we end up with zero surface gravities which is irrelevant for our case of study.

Taking two roots very close ($r_{i}=r_{j}+\varepsilon $) leads to the coalescence of these singularities in the limit\ $\varepsilon \rightarrow 0$. We then have three regular singular points and one irregular singular point and the equation is no longer Fuchsian. For example taking $r_{2}=r_{1}+\varepsilon $, in the limit $\varepsilon \rightarrow 0$ we have regular singular points at $\{r_{3},r_{4},\infty \}$ and an irregular singular point at $\{r_{1}\}$. We can coalesce more than one point by taking other roots at close values. However, the coalescence of singular points does not help us finding a solution to our equation.
%%%%%%%%%%%%%%%%%%%%%%%%%%%%%%%%%%%%%%%%%%%%%%%
\subsection{Near-horizon (near-)extremal cases}
%%%%%%%%%%%%%%%%%%%%%%%%%%%%%%%%%%%%%%%%%%%%%%%
We can obtain the near-horizon solution using the scaling transformations
\begin{eqnarray}
r &=&r_{1}(1+\lambda \rho ), \\
\phi &=&\psi +\Omega _{1}t, \\
t &=&\frac{\tau }{2\pi r_{1}\frac{dT_{1}}{dr_{1}}\lambda },
\end{eqnarray}
defined in \cite{Chow:2008dp}. Here $r_{1}$ is the largest root of the horizon equation and $\rho$ is the new radial coordinate. For the near-extremal limit we need
\begin{equation}
r_{2}=r_{1}(1+p\lambda ),
\end{equation}
and $p=0$ is the extremal limit \cite{Birkandan:2011fr}. Defining
\begin{eqnarray}
W_{1} &\equiv &W(r_{1}, y )=r_{1}^{2}+y-v^{2}, \\
L_{1} &\equiv &L(r_{1})=r_{1}^{2}+a^{2}-v^{2}, \\
K_{1} &\equiv &{g}^{2}\left( r_{1}-r_{3}\right) \left( r_{1}-r_{4}\right) ,
\end{eqnarray}
we get
\begin{eqnarray}
ds^{2}&=&-{\frac{4W_{1}\rho \left( \rho -p\right) }{K_{1}}d}\tau ^{2}+{\frac{%
\left( {a}^{2}-y\right) \Theta _{g}\left( {K}_{1}L_{1}d\psi +4\Xi ar_{1}\rho
d\tau \right) ^{2}}{{a}^{2}\Xi ^{2}K_{1}^{2}W_{1}}} \notag \\
&&+W_{1}\left[ {\frac{d\rho^{2}}{K_{1}\rho \left( \rho -p\right) }}+{\frac{{dy}^{2}}{4y\left( {a}^{2}-y\right) \Theta _{g}}}\right] {,}
\end{eqnarray}
as the near-horizon near-extremal (nhne) metric in the limit $\lambda\rightarrow 0$. The Frolov-Thorne temperature can be found as \cite{Chow:2008dp}
\begin{equation}
T_{\psi }=-\frac{\frac{dT}{dr_{1}}}{\frac{d\Omega _{1}}{dr_{1}}}=\frac{K_{1}L_{1}}{8\pi a\Xi r_{1}}.
\end{equation}
We can use a similar Ansatz, namely
\begin{equation}
\varphi _{nhne}=e^{-i\omega \tau }e^{ik\psi }X_{nhne}(\rho )Y_{nhne}(y),
\end{equation}
in the massless Klein-Gordon equation and the angular part is still in the form of the general Heun's equation. The radial part reads
\begin{equation}
\frac{d^{2}X_{nhne}}{d\rho ^{2}}+\frac{2\left( \rho -\frac{p}{2}\right) }{%
\rho \left( \rho -p\right) }\frac{dX_{nhne}}{d\rho }+\left\{ \left[ \frac{%
k\rho +2\pi T_{\psi }\omega }{4\pi T_{\psi }\rho \left( \rho -p\right) }%
\right] ^{2}+\frac{c_{2}}{32a^{2}\pi ^{2}\Xi ^{2}r_{1}^{2}T_{\psi }^{2}\rho
\left( \rho -p\right) }\right\} X_{nhne}=0,
\end{equation}
where $c_{2}$ is the separation constant. The solution can be found in terms of the hypergeometric functions
\begin{eqnarray}
\fl X_{nhne} &=&C_{1}{\rho }^{{\frac{-i}{2\omega p}}}\left( \rho -p\right) ^{{%
\frac{i\left( 2\pi \omega T_{\psi }+kp\right) }{4\pi pT_{\psi }}}}{F}\left( {%
{\frac{-U_{1}+U_{2}}{U_{4}}},\,{\frac{U_{1}+U_{2}}{U_{4}}};\,{\frac{%
p-i\omega }{p}};\,{\frac{\rho }{p}}}\right)  \notag \\
\fl &&+C_{2}{\rho }^{{\frac{i}{2\omega p}}}\left( \rho -p\right) ^{{\frac{%
i\left( 2\pi \omega T_{\psi }+kp\right) }{4\pi pT_{\psi }}}}{F}\left( {{%
\frac{U_{1}p+U_{3}}{U_{4}p}},{\frac{-U_{1}p+U_{3}}{U_{4}p}};\,{\frac{%
p+i\omega }{p}};\,{\frac{\rho }{p}}}\right) ,
\end{eqnarray}
where
\begin{eqnarray}
U_{1} &=&\sqrt{2\left[ \left( 8\,{\pi }^{2}T_{\psi }^{2}-2\,{k}^{2}\right) {a%
}^{2}\Xi ^{2}r_{1}^{2}-c_{2}\right] }, \\
U_{2} &=&2\,\Xi \,ar_{1}\,\left( ik+2\pi T_{\psi }\right) , \\
U_{3} &=&8\pi r_{1}\,a\Xi \,\left[ \left( i\omega +\frac{p}{2}\right)
T_{\psi }+\frac{ikp}{4\pi }\right] , \\
U_{4} &=&8\pi T_{\psi }\,\Xi \,ar_{1}.
\end{eqnarray}
In the near horizon extremal (nhe) limit ($p\rightarrow 0$), we have the Whittaker functions as radial solutions
\begin{equation}
X_{nhe}=C_{1}\,M{\left( -{\frac{ik}{4\pi T_{\psi }}},{\frac{U_{1}}{U_{4}},%
\frac{i\omega }{\rho }}\right) }+C_{2}\,W{\left( -{\frac{ik}{4\pi T_{\psi }}}%
,{\frac{U_{1}}{U_{4}},\frac{i\omega }{\rho }}\right) ,}
\end{equation}
which are also hypergeometric-type functions. Here, we displayed the $M_{\mu_{1},\mu _{2}}(z)$and $W_{\mu _{1},\mu _{2}}(z)$ functions as $M(\mu_{1},\mu _{2},z)$ and $W(\mu _{1},\mu _{2},z)$ respectively. Our choice is different from the common literature (e.g., see \cite{bateman}) but it shows the arguments in a clear way.

The emergence of hypergeometric functions signifies the SL(2,\textbf{R})$^{2} $ invariance of conformal symmetry as they form representations of SL(2,\textbf{R}).

Following \cite{Becker:2010jj,Becker:2010dm}, we can write the retarded Green's function for the near horizon near-extremal case. Using the asymptotic expansion ($\rho \to \infty$) of the solution we have
\begin{equation}
X_{nhne}\sim A\rho ^{-\Delta _{-}}+B\rho ^{-\Delta _{+}},
\end{equation}
where
\begin{eqnarray}
\Delta _{-} &=&\frac{U_{2}-U_{1}}{U_{4}}, \\
\Delta _{+} &=&\frac{U_{2}+U_{1}}{U_{4}},
\end{eqnarray}
and we get
\begin{equation}
G_{R_{nhne}}\sim \frac{B}{A}=\frac{\left( -\frac{1}{p}\right) ^{-\frac{2U_{1}%
}{U_{4}}}\Gamma \left( -\frac{2U_{1}}{U_{4}}\right) \Gamma \left( \frac{%
U_{1}+U_{2}}{U_{4}}\right) \Gamma \left( \frac{U_{1}-U_{2}+U_{4}}{U_{4}}-%
\frac{i\omega }{p}\right) }{\Gamma \left( \frac{2U_{1}}{U_{4}}\right) \Gamma
\left( \frac{U_{2}-U_{1}}{U_{4}}\right) \Gamma \left( -\frac{%
U_{1}+U_{2}-U_{4}}{U_{4}}-\frac{i\omega }{p}\right) }.
\end{equation}
Similarly, for the extremal case we have
\begin{equation}
X_{nhe}\sim A\rho ^{-\frac{1}{2}-\beta }+B\rho ^{-\frac{1}{2}+\beta },
\end{equation}
where
\begin{equation}
\beta =\frac{U_{1}}{U_{4}},
\end{equation}
and the retarded Green's function reads
\begin{equation}
G_{R_{nhe}}\sim \frac{B}{A}=\frac{(i\omega )^{\frac{2U_{1}}{U_{4}}}\Gamma
\left( -\frac{2U_{1}}{U_{4}}\right) \Gamma \left( \frac{ik}{4\pi T_{\psi }}+%
\frac{U_{1}}{U_{4}}+\frac{1}{2}\right) }{\Gamma \left( \frac{2U_{1}}{U_{4}}%
\right) \Gamma \left( \frac{ik}{4\pi T_{\psi }}-\frac{U_{1}}{U_{4}}+\frac{1}{%
2}\right) }.
\end{equation}
This result can be regarded as the starting point of the correlation function calculations of the boundary conformal field theory operators \cite{Son:2002sd,Herzog:2002pc}. Note that as $p\rightarrow 0$, we have $G_{R_{nhne}}\rightarrow G_{R_{nhe}}$ up to a factor as the last terms in the numerator and denominator cancel each other in $G_{R_{nhne}}$.
%%%%%%%%%%%%%%%%%%%%%%%%%%%%%%%%%%%%%%%%%%%%%%%
\section{Conclusions}
%%%%%%%%%%%%%%%%%%%%%%%%%%%%%%%%%%%%%%%%%%%%%%%
We studied  the massless Klein-Gordon equation in the background of the most general black hole in  four-dimensional $\mathcal{N}=2$ gauged supergravity with $U(1)^2$ gauge symmetry, found in \cite{Chow:2013gba}. It is a rotating  AdS black hole with two independent electric charges and  two independent magnetic charges. Due to $S$-duality,  its  metric is of the same structure as that \cite{Chong:2004na} of the black hole specified by two electric charges, only. The existence of a Killing-Yano tensor with torsion ensures the separability of the wave equation \cite{Chow:2013gba}. The angular part, which is a Fuchsian equation with four regular singularities could be solved in terms of the general Heun's functions. The radial part of the scalar wave equation is another Fuchsian equation with five regular singularities at the four poles of the horizon equation and at positive infinity.

We focused on the analysis of the radial part of the equation,  and showed explicitly that the  poles, associated with each  black hole horizon,  have residua specified by the specific combination of the surface gravity and angular velocity at the respective horizon. The analysis confirms the  analogous anticipated structure obtained \cite{Birkandan:2014vga} for the most general black holes in five-dimensional maximally supersymmetric gauge supergravity \cite{Wu:2011gq}.  The analysis of the radial part  of the wave equation therefore carries the key information about  the internal structure of the black hole, associated with all the horizons.

We also analyzed the near-horizon (near-)extremal cases of the solution. As anticipated, in  these limits  we obtained the radial wave equation whose solutions are hypergeometric functions, thus revealing a hidden conformal symmetry,  with an underlying SL(2,\textbf{R}) conformal invariance.  This enabled us to calculate the retarded Green's functions, constituting a starting point for correlation function calculations of the boundary conformal field theory operators.
%%%%%%%%%%%%%%%%%%%%%%%%%%%%%%%%%%%%%%%%%%%%%%%
\ack
%%%%%%%%%%%%%%%%%%%%%%%%%%%%%%%%%%%%%%%%%%%%%%%
We would like to thank Sera Cremonini for discussions. The research of TB is supported by TUBITAK, the Scientific and Technological Council of Turkey and Istanbul Technical University [\.{I}T\"{U} BAP 37519]. MC would like to thank the Department of Physics at the Istanbul Technical University for hospitality during the early course of this work. MC's visit to Istanbul Technical University was supported by TUBITAK \textquotedblleft 2221-Fellowship for Visiting Scientists" with number 1059B211400897. MC is also supported by the DOE Grant DOE-EY-76-02-3071, the Fay R. and Eugene L. Langberg Endowed Chair, the Slovenian Research Agency (ARRS), and the Simons Foundation Fellowship No. 267489.
%%%%%%%%%%%%%%%%%%%%%%%%%%%%%%%%%%%%%%%%%%%%%%%
\appendix
\section{Charge-dependence of the metric parameters}
%%%%%%%%%%%%%%%%%%%%%%%%%%%%%%%%%%%%%%%%%%%%%%%
We will study the charge dependence of the metric parameters explicitly in this Appendix. We will give the main definitions and the reader should refer to the Section IV of Chow and Comp\`{e}re's paper for details \cite{Chow:2013gba}, where all the gauge potentials and axio-dilaton fields are also presented. Here we focus only on the functions appearing in the metric, given as:
\begin{eqnarray}
R_{g}&=&r^{2}-2mr+a^{2}+e^{2}-N_{g}^{2} \notag \\
&&+g^{2}\left[r^{4}+(a^{2}+6N_{g}^{2}-2v^{2})r^{2}+3N_{g}^{2}\left( a^{2}-N_{g}^{2}\right) \right] , \\
L&=&r^{2}+(N_{g}+a)^{2}-v^{2}, \\
\Theta _{g}&=&1-a^{2}g^{2}\cos ^{2}\theta -4ag^{2}N_{g}\cos \theta, \\
W&=&r^{2}+(N_{g}+a\cos \theta )^{2}-v^{2}, \\
\Xi &=&1-a^{2}g^{2}-4aN_{g}g^{2},
\end{eqnarray}
and the two charge-dependent parts are given by
\begin{eqnarray}
v^{2} &=&\beta ^{-2}(\Delta _{\Delta r}^{2}+\Delta _{\Delta u}^{2}), \\
e^{2} &=&\beta ^{-4} \{ 2\left( \widehat{m}\Sigma _{\Delta r}+\widehat{n}\Sigma _{\Delta u}\right) +\Sigma _{\Delta r}^{2}+\Sigma _{\Delta u}^{2} \notag \\
&&-g^{2}\left[ \Delta _{\Delta r}^{2}(\widehat{a}^{2}-\Delta _{\Delta r}^{2})-\Delta _{\Delta u}^{2}(\widehat{a}^{2}+\Delta _{\Delta u}^{2})\right] \} .
\end{eqnarray}
Here
\begin{eqnarray}
\Sigma _{\Delta r} &=&\tfrac{1}{2}(\Delta \widehat{r}_{1}+\Delta \widehat{r}%
_{2}), \\
\Delta _{\Delta r} &=&\tfrac{1}{2}(\Delta \widehat{r}_{2}-\Delta \widehat{r}%
_{1}), \\
\Sigma _{\Delta u} &=&\tfrac{1}{2}(\Delta \widehat{u}_{1}+\Delta \widehat{u}%
_{2}), \\
\Delta _{\Delta u} &=&\tfrac{1}{2}(\Delta \widehat{u}_{2}-\Delta \widehat{u}%
_{1}),
\end{eqnarray}
where
\begin{eqnarray}
\Delta \widehat{r}_{1} &=&\widehat{m}[\cosh (2\delta _{1})\cosh (2\gamma
_{2})-1]+\widehat{n}\sinh (2\delta _{1})\sinh (2\gamma _{1}), \\
\Delta \widehat{r}_{2} &=&\widehat{m}[\cosh (2\delta _{2})\cosh (2\gamma
_{1})-1]+\widehat{n}\sinh (2\delta _{2})\sinh (2\gamma _{2}), \\
\Delta \widehat{u}_{1} &=&\widehat{n}[\cosh (2\delta _{1})\cosh (2\gamma
_{2})-1]-\widehat{m}\sinh (2\delta _{1})\sinh (2\gamma _{1}), \\
\Delta \widehat{u}_{2} &=&\widehat{n}[\cosh (2\delta _{2})\cosh (2\gamma
_{1})-1]-\widehat{m}\sinh (2\delta _{2})\sinh (2\gamma _{2}).
\end{eqnarray}
Here, $\widehat{m}$ is the mass-related term and $\widehat{n}~$is the NUT charge-related term. For the asymptotically flat case we have
\begin{eqnarray}
M &=&\widehat{m}+\Sigma _{\Delta r}, \\
N &=&\widehat{n}+\Sigma _{\Delta u},
\end{eqnarray}
where $M$ is the physical mass and $N$ is the NUT charge. Then one can define the electric and magnetic charges by
\begin{eqnarray}
Q_{1} &=&\frac{\partial M}{\partial \delta _{1}},Q_{2}=\frac{\partial M}{%
\partial \delta _{2}}, \\
P_{1} &=&-\frac{\partial N}{\partial \delta _{1}},P_{2}=-\frac{\partial N}{%
\partial \delta _{2}}.
\end{eqnarray}
For the gauged case, we have
\begin{eqnarray}
\beta ^{2} =\frac{1+g^{2}(\widehat{a}^{2}+2\Delta _{\Delta u}^{2})}{1+g^{2}(a^{2}+6N_{g}^{2})}, \\
\frac{a^{2}}{(1+a^{2}g^{2})^{2}} =\frac{\widehat{a}^{2}-\Sigma _{\Delta u}^{2}+g^{2}\Delta _{\Delta u}^{2}(\widehat{a}^{2}+\Delta _{\Delta u}^{2})}{[1+g^{2}(\widehat{a}^{2}+2\Delta _{\Delta u}^{2})]^{2}}.
\end{eqnarray}
As one can see from the following special charge relations, the form of the solution is preserved and it is no more complicated than that of the special cases. The two charge dependent parts, $v^{2}$ and $e^{2}$, can take specific values but they do not change their structures.
%%%%%%%%%%%%%%%%%%%%%%%%%%%%%%%%%%%%%%%%%%%%%%%
\subsection{Pairwise equal electric and magnetic charges}
%%%%%%%%%%%%%%%%%%%%%%%%%%%%%%%%%%%%%%%%%%%%%%%
This is the case of Einstein-Maxwell-de Sitter gravity theory, which was analyzed as a special case in  \cite{Chow:2013gba}. For pairwise equal electric and magnetic charges ($\delta _{1}=\delta _{2}$ and $\gamma _{1}=\gamma _{2}$) we have
\begin{eqnarray}
\Delta \widehat{r}_{1} &=&\widehat{m}[\cosh (2\delta _{1})\cosh (2\gamma
_{1})-1]+\widehat{n}\sinh (2\delta _{1})\sinh (2\gamma _{1}), \\
\Delta \widehat{u}_{1} &=&\widehat{n}[\cosh (2\delta _{1})\cosh (2\gamma
_{1})-1]-\widehat{m}\sinh (2\delta _{1})\sinh (2\gamma _{1}), \\
\Delta \widehat{r}_{2} &=&\Delta \widehat{r}_{1},\Delta \widehat{u}%
_{2}=\Delta \widehat{u}_{1}.
\end{eqnarray}
Then we have
\begin{eqnarray}
\Sigma _{\Delta r} &=&\Delta \widehat{r}_{1},\text{ \ \ \ \ }\Sigma _{\Delta
u}=\Delta \widehat{u}_{1}, \\
\Delta _{\Delta r} &=&0,\text{ \ \ \ }\Delta _{\Delta u}=0.
\end{eqnarray}
For the metric parameters we have
\begin{eqnarray}
v^{2} &=&0, \\
e^{2} &=&\beta ^{-4}\left[ 2\left( \widehat{m}\Sigma _{\Delta r}+\widehat{n}%
\Sigma _{\Delta u}\right) +\Sigma _{\Delta r}^{2}+\Sigma _{\Delta u}^{2}%
\right] .
\end{eqnarray}
It was shown in the original article that
\begin{equation}
e^{2}=\frac{1}{\beta ^{4}}\left( Q_{1}^{2}+P_{1}^{2}\right) ,
\end{equation}
in terms of the physical electromagnetic charges.
%%%%%%%%%%%%%%%%%%%%%%%%%%%%%%%%%%%%%%%%%%%%%%%
\subsection{Zero magnetic charges}
%%%%%%%%%%%%%%%%%%%%%%%%%%%%%%%%%%%%%%%%%%%%%%%
In the case of the zero magnetic charges ($\gamma _{1}=\gamma _{2}=0$), the solution was found  in \cite{Chong:2004na}. The NUT charge can be canceled by setting $\widehat{n}=0$. We have $\beta =1$ and $\widehat{a}=a$.  We obtain
\begin{eqnarray}
\Delta \widehat{r}_{1} &=&2\widehat{m}s_{1}^{2},\text{ \ \ \ \ }\Delta
\widehat{r}_{2}=2\widehat{m}s_{2}^{2}, \\
\Delta \widehat{u}_{1} &=&0,\text{ \ \ \ \ }\Delta \widehat{u}_{2}=0,
\end{eqnarray}
where $s_{i}=\sinh (\delta _{i})$. Then we have
\begin{eqnarray}
\Sigma _{\Delta r} &=&\widehat{m}\left( s_{1}^{2}+s_{2}^{2}\right) ,\text{ \
\ \ \ }\Delta _{\Delta r}=\widehat{m}\left( s_{1}^{2}-s_{2}^{2}\right) , \\
\Sigma _{\Delta u} &=&0,\text{ \ \ \ \ }\Delta _{\Delta u}=0.
\end{eqnarray}
We finally get
\begin{eqnarray}
v^{2} &=&\Delta _{\Delta r}^{2}, \\
e^{2} &=&2\widehat{m}\Sigma _{\Delta r}+\Sigma _{\Delta r}^{2}-g^{2}\Delta
_{\Delta r}^{2}(a^{2}-\Delta _{\Delta r}^{2}),
\end{eqnarray}
for the two charge-dependent parts.

Comparing the results with \cite{Chong:2004na}, one can see that the structure is the same due to the $S$-duality of the theory.
%%%%%%%%%%%%%%%%%%%%%%%%%%%%%%%%%%%%%%%%%%%%%%%
\section{Special limits of the radial equation}
%%%%%%%%%%%%%%%%%%%%%%%%%%%%%%%%%%%%%%%%%%%%%%%
We present some possible special limits of the radial equation in this Appendix.

The full radial equation given by equation (\ref{fullradial}) has no limiting cases for the general metric we have been studying. It is possible to have some subcases of the metric, and they have been studied in the literature extensively (see \cite{Chow:2013gba} for these subcases).

We have in the near-horizon near-extremal limit
\begin{equation}
\frac{d^{2}X_{nhne}}{d\rho ^{2}}+\frac{2\left( \rho -\frac{p}{2}\right) }{%
\rho \left( \rho -p\right) }\frac{dX_{nhne}}{d\rho }+\left\{ \left[ \frac{%
k\rho +2\pi T_{\psi }\omega }{4\pi T_{\psi }\rho \left( \rho -p\right) }%
\right] ^{2}+\frac{c_{2}}{32a^{2}\pi ^{2}\Xi ^{2}r_{1}^{2}T_{\psi }^{2}\rho
\left( \rho -p\right) }\right\} X_{nhne}=0,
\end{equation}
for the radial equation, as found in the text. Here, $p$ is the extremality parameter. The solution of this equation can be given in terms of the hypergeometric functions. As a following limiting case, we can immediately write the extremal case ($p\rightarrow 0$), which was also obtained in the text. The radial equation corresponding to this case is the Whittaker equation with the solution
\begin{equation}
X_{nhe}=C_{1}\,M{\left( -{\frac{ik}{4\pi T_{\psi }}},{\frac{U_{1}}{U_{4}},%
\frac{i\omega }{\rho }}\right) }+C_{2}\,W{\left( -{\frac{ik}{4\pi T_{\psi }}}%
,{\frac{U_{1}}{U_{4}},\frac{i\omega }{\rho }}\right) ,}
\end{equation}
and this equation is a modified form of the confluent hypergeometric equation \cite{nist}. The Whittaker functions have relations with Bessel functions if they have zero first arguments. In our case, this means $-{\frac{ik}{4\pi T_{\psi }}\rightarrow 0}$. The easiest approach would be setting the azimuthal eigenvalue $k$ to zero. Then remembering
\begin{eqnarray}
U_{1} &=&\sqrt{2\left[ \left( 8\,{\pi }^{2}T_{\psi }^{2}-2\,{k}^{2}\right) {a%
}^{2}\Xi ^{2}r_{1}^{2}-c_{2}\right] }, \\
U_{4} &=&8\pi T_{\psi }\,\Xi \,ar_{1},
\end{eqnarray}
we have
\begin{equation}
X_{nhe,\{k=0\}}={\frac{1}{\sqrt{\rho }}\left[ C_{1}\,J{\left( -\,{\frac{U_{1}%
}{U_{4}},\frac{\omega }{2\rho }}\right) +}C_{2}\,Y{\left( -{\frac{U_{1}}{%
U_{4}}},{\frac{\omega }{2\rho }}\right) }\right] ,}
\end{equation}
in terms of Bessel functions of the first and second kind. The other approach, which gives the same result, would be sending the Frolov-Thorne temperature $T_{\psi }$ to infinity which is unphysical. Furthermore, if we have $\frac{U_{1}}{U_{4}}=\frac{1}{2}$ (i.e. $c_{2}=0$), in addition to $k=0$, we obtain
\begin{equation}
X_{nhe,\{k=0,c_{2}=0\}}=C_{1}e^{\frac{i\omega }{2\rho }}{+}C_{2}e^{-\frac{%
i\omega }{2\rho }},
\end{equation}
as a much simpler form.

For a nonzero $p$ and using $\{k=0,$ $c_{2}=0\}$ we get
\begin{equation}
X_{nhne,\{k=0,c_{2}=0\}}=C_{1}\sin \left( \,{\frac{\omega \left[ \ln \rho
-\ln \left( \rho -p\right) \right] }{2p}}\right) +C_{2}\cos \left( \,{\frac{%
\omega \left[ \ln \rho -\ln \left( \rho -p\right) \right] }{2p}}\right) ,
\end{equation}
as a limiting case for the near-extremal limit. We need both the $k=0$ and $c_{2}=0$ limits to have reductions from the hypergeometric equation to simpler forms.
%%%%%%%%%%%%%%%%%%%%%%%%%%%%%%%%%%%%%%%%%%%%%%%
\section*{References}
%%%%%%%%%%%%%%%%%%%%%%%%%%%%%%%%%%%%%%%%%%%%%%%

\end{document}